\documentclass[12pt]{article}

\usepackage{amssymb}
\usepackage{amsmath}
\usepackage{amsfonts}

\usepackage{graphicx}
\usepackage{epstopdf} 
\usepackage{amsthm}
\usepackage{enumerate}

\newcommand{\blind}{0}
\newcommand{\bfrA}{\mathbf{A}}
\newcommand{\bfrP}{\mathbf{P}}
\newcommand{\bfrX}{\mathbf{X}}
\newcommand{\bfrD}{\mathbf{D}}
\newcommand{\Deff} {D_{\mbox{\textit{\scriptsize{eff}}}}}

\newcommand{\Fcc}[2]{\ensuremath{Fc_{#1} \!\! < \!\! c_{#2}}}

\mathchardef\mhyphen="2D
\newcommand{\OofAOA}{\mbox{\textit{OofA}}\mhyphen\mbox{OA}}

\newcommand{\ssection}[1]{%
  \section[#1]{\centering\normalfont\scshape #1}}

\newtheorem{theorem0}{Theorem}
\newtheorem{definition0}{Definition}
\newtheorem{proposition0}{Proposition}

\begin{document}

\def\spacingset#1{\renewcommand{\baselinestretch}%
{#1}\small\normalsize} \spacingset{1}



\if0\blind
{
  \title{\bf The Design of Order-of-Addition Experiments}
  \author{JOSEPH G. VOELKEL\\\textit{School of Mathematical Sciences}\\ \textit{Rochester Institute of Technology, Rochester, NY 14623}}
  \maketitle
} \fi

\if1\blind
{
  \bigskip
  \bigskip
  \bigskip
  \begin{center}
    {\LARGE\bf The Design of Order-of-Addition Experiments}
\end{center}
  \medskip
} \fi

\bigskip
\begin{abstract}
We introduce systematic methods to create optimal designs for order-of-addition (OofA) experiments, those that study the order in which $m$ components are applied---for example, the order in which chemicals are added to a reaction or layers are added to a film. Full designs require $m!$ runs, so we investigate design fractions. Balance criteria for creating such designs employ an extension of orthogonal arrays (OA's) to OofA-OA's. A connection is made between $D$-efficient and OofA-OA designs. Necessary conditions are found for the number of runs needed to create OofA-OA's of strengths 2 and 3. We create a number of new, optimal, designs: 12-run OofA-OA's in 4 and 5 components, 24-run OofA-OA's in 5 and 6 components, and near OofA-OA's in 7 components. We extend these designs to include (a) process factors, and (b) the common case in which component orderings are restricted. We also suggest how such designs may be analyzed.
\end{abstract}

\noindent%

\newpage

\ssection{INTRODUCTION}\label{sec:Int}

An early reference to an order-of-addition (OofA) experiment (Fisher(1971)) is that of a lady tasting tea, who claimed she could distinguish whether the tea or the milk was first added to her cup. Fisher's experiment consisted of four replications of each of tea$\rightarrow$milk and milk$\rightarrow$tea.

More complex OofA applications appear in a number of areas, including chemical, food, and film, although in the author's experience such applications tend to be proprietary. An application in the public arena includes Preuss, et al.\ (2009, p. 24), who note that ``The next step [in an investigation] is usually to determine the order of action of the protein with respect to known treatments required at the same general transport step. This approach is sometime referred to an order of addition experiment...'' A PerkinElmer (2016) reference states ``Order of addition can influence the signal generated to a large extent. \emph{The optimal order in which assay components interact should always be determined empirically}... some binding partners may interfere with the association of other binding partners if allowed to interact in the wrong order.''(italics ours.) Other references to OofA experiments include Black, et al.\ (2001) and Kim, et al.\ (2001). None of these references mention a strategy of experimenter design---rather, a small subset of select orders was investigated.

In this paper, we will denote each material whose order of addition is being considered as a \emph{component}. If there are $m$ components, say $c_0,\ldots,c_{m-1}$, then there are $m!$ possible combinations of orders. An experiment with $5$ components or more would normally be considered excessively large, so our interest is to find good fractions of such designs. Our objectives in this paper are to:
\begin{enumerate}
\item Review past work in the field;
\item Consider goodness criteria for fractionating an OofA design;
\item Investigate the goodness of the designs created by these criteria, including a comparison of these designs to those from past work;
\item Extend these designs to include process factors as well---for example, varying the temperature or the amount of $c_1$;
\item Consider designs in which restrictions are placed on component ordering.
\end{enumerate}

\ssection{PAST WORK}\label{sec:PastWork}

Given the importance of order of addition in certain experimental areas, it is surprising that little research has been done to create good OofA designs. The only direct statistical reference for creating such designs appears to be Van Nostrand (1995), whose primary approach was to select a fraction of such combinations with good properties, as follows:

\begin{enumerate}
\item Create ``pseudo factors'' (which in this paper we will call \emph{pair-wise ordering} (PWO) factors). There are ${m\choose 2}$ PWO factors, corresponding to all pairs of component orders $\{\Fcc{k}{l}, 0\le k<l \le m-1\}$, where $F$ is included in these names to emphasize these are factors, and each $\Fcc{k}{l}$ is at two levels, $1$ and $0$, to indicate whether or not component $k$ is added before component $l$, respectively. So, if $m=4$, the six PWO factors are $\Fcc{0}{1},\;\Fcc{0}{2},\;\Fcc{0}{3}\;\Fcc{1}{2},\;\Fcc{1}{3},\;\Fcc{2}{3}$---see Table~\ref{tab:OofA4_24} for an example of a design in all $m!$ runs. (The $\bfrD_F$ and $\bfrP_F$ notation will be explained in the next section.)
\item Create a tentative two-level design in these ${m\choose 2}$ PWO factors, fractionated to a suggested number of starting runs. For $m=5$, for example, with $10$ PWO factors, this might correspond to a $2^{10-3}$ design, the hope being that, because the full $2^{10}$ would include all valid $5!=120$ combinations, then a particular $2^{10-3}$ design might include about $120(2^{-3})=15$ valid combinations.
\item Reduce the number of runs in the fraction to include only valid combinations. For example, if the fraction includes a suggested run such as \mbox{$c_0<c_1,\;c_0>c_2,\;c_1<c_2$,} such a run would be excluded.
\end{enumerate}

\begin{table}[!htbp] \centering
\caption{Four-factor OofA design $\bfrD_F$ in 24 runs, with PWO design matrix $\bfrP_F$.}
\label{tab:OofA4_24}
\small
\begin{tabular}{@{\extracolsep{-10pt}} rcccccccccccc} 
\\
\hline
\hline
\\
    & \hspace{.2in}  &     \multicolumn{4}{c}{Design $\bfrD_F$}  & & \multicolumn{6}{c}{PWO Design Matrix $\bfrP_F$} \\
\cline{8-13}
&   &     \multicolumn{4}{c}{}  & & $\Fcc{0}{1}$ & $\Fcc{0}{2}$ & $\Fcc{0}{3}$ & $\Fcc{1}{2}$ & $\Fcc{1}{3}$ & $\Fcc{2}{3}$ \\\cline{3-6} \cline{8-13}
$1$ &   &\hspace{.05in}$0$\hspace{.05in}& \hspace{.05in}$1$\hspace{.05in} & \hspace{.05in}$2$\hspace{.05in} & \hspace{.05in}$3$\hspace{.05in} &   & $1$ & $1$ & $1$ & $1$ & $1$ & $1$ \\
$2$ &   & $0$ & $1$ & $3$ & $2$ &\hspace{.2in}   & $1$ & $1$ & $1$ & $1$ & $1$ & $0$ \\
$3$ &   & $0$ & $2$ & $1$ & $3$ &   & $1$ & $1$ & $1$ & $0$ & $1$ & $1$ \\
$4$ &   & $0$ & $2$ & $3$ & $1$ &   & $1$ & $1$ & $1$ & $0$ & $0$ & $1$ \\
$5$ &   & $0$ & $3$ & $1$ & $2$ &   & $1$ & $1$ & $1$ & $1$ & $0$ & $0$ \\
$6$ &   & $0$ & $3$ & $2$ & $1$ &   & $1$ & $1$ & $1$ & $0$ & $0$ & $0$ \\
$7$ &   & $1$ & $0$ & $2$ & $3$ &   & $0$ & $1$ & $1$ & $1$ & $1$ & $1$ \\
$8$ &   & $1$ & $0$ & $3$ & $2$ &   & $0$ & $1$ & $1$ & $1$ & $1$ & $0$ \\
$9$ &   & $1$ & $2$ & $0$ & $3$ &   & $0$ & $0$ & $1$ & $1$ & $1$ & $1$ \\
$10$ &   & $1$ & $2$ & $3$ & $0$ &   & $0$ & $0$ & $0$ & $1$ & $1$ & $1$ \\
$11$ &   & $1$ & $3$ & $0$ & $2$ &   & $0$ & $1$ & $0$ & $1$ & $1$ & $0$ \\
$12$ &   & $1$ & $3$ & $2$ & $0$ &   & $0$ & $0$ & $0$ & $1$ & $1$ & $0$ \\
$13$ &   & $2$ & $0$ & $1$ & $3$ &   & $1$ & $0$ & $1$ & $0$ & $1$ & $1$ \\
$14$ &   & $2$ & $0$ & $3$ & $1$ &   & $1$ & $0$ & $1$ & $0$ & $0$ & $1$ \\
$15$ &   & $2$ & $1$ & $0$ & $3$ &   & $0$ & $0$ & $1$ & $0$ & $1$ & $1$ \\
$16$ &   & $2$ & $1$ & $3$ & $0$ &   & $0$ & $0$ & $0$ & $0$ & $1$ & $1$ \\
$17$ &   & $2$ & $3$ & $0$ & $1$ &   & $1$ & $0$ & $0$ & $0$ & $0$ & $1$ \\
$18$ &   & $2$ & $3$ & $1$ & $0$ &   & $0$ & $0$ & $0$ & $0$ & $0$ & $1$ \\
$19$ &   & $3$ & $0$ & $1$ & $2$ &   & $1$ & $1$ & $0$ & $1$ & $0$ & $0$ \\
$20$ &   & $3$ & $0$ & $2$ & $1$ &   & $1$ & $1$ & $0$ & $0$ & $0$ & $0$ \\
$21$ &   & $3$ & $1$ & $0$ & $2$ &   & $0$ & $1$ & $0$ & $1$ & $0$ & $0$ \\
$22$ &   & $3$ & $1$ & $2$ & $0$ &   & $0$ & $0$ & $0$ & $1$ & $0$ & $0$ \\
$23$ &   & $3$ & $2$ & $0$ & $1$ &   & $1$ & $0$ & $0$ & $0$ & $0$ & $0$ \\
$24$ &   & $3$ & $2$ & $1$ & $0$ &   & $0$ & $0$ & $0$ & $0$ & $0$ & $0$ \\
\hline \\
\end{tabular}
\end{table}

For five factors, Van Nostrand examined the eight fractions from a particular $2^{10-3}$ family---that is, all eight combinations of $\pm G_1,\;\pm G_2, \; \pm G_3$, where the $G_i$'s are the words in the generator. For each combination, he considered the main-effects model, and then used the mean VIF's (variance-inflation factors) from each model to measure the goodness of each design. In his example, the number of valid runs from his 128-run fractions ranged from 13 to 17; the 13-run designs were singular, and the other designs had mean VIF's in the range from 3.3 to 3.6. (The mean VIF in the full, 120-run, design is 2.) He noted that different sets of words in the generator lead to different designs, and that ``[p]rinciples for assigning letters to generators to produce designs with low variance inflation factors are not yet apparent.'' Also note that even if a full-factorial design is considered for $m$ component, only the fraction $m!/2^{m(m-1)/2}$ of these runs are valid. In Van Nostrand's 5-component example, only about 12\% of the runs from the full-factorial design would be valid, which is consistent with the fractions valid in his fractional-factorial designs. This fraction drops quickly with $m$---even for $m=6$, the fraction is only 2\%.

An author whose designs appear to be more widely used for OofA experiments is Williams (1949, 1950). However,  Williams' intent was to create designs not for OofA experiments, but for crossover experiments---experiments in which each subject receives $m$ treatments in succession. His designs address the possibility that one or more preceding treatments as well as the current one may affect the response---his designs are balanced for these so-called residual effects. Williams showed that designs that are balanced for the effect of only one preceding treatment exist for $m$ treatments in one $m\times m$ Latin Square ($m$ runs) if $m$ is even---this is known as a cyclic row-complete Latin Square---and two such Latin Squares ($2m$ runs) if $m$ is odd. In these Latin Squares, rows indicate the run (or subject) number, columns the treatment order, and Latin letters the treatment.

In addition, if $m$ is a prime or a power of a prime, then Williams showed that a design in $m-1$ Latin Squares ($m(m-1)$ runs) may be created that is balanced for any number of preceding treatments, where their interactions are not taken into account. Williams also created designs for $m$ treatments, again in $m-1$ Latin Squares, that are balanced for each pair of preceding treatments in both of their orders, so that residual-effect interactions may be taken into account.

\ssection{APPROACH AND CRITERIA}\label{sec:Crit}

Williams' designs lead to certain balance properties. However, these properties are not directly connected to OofA considerations. For example, the natural factors in Williams' designs, as shown in his 1949 example, are replications, order, treatments (direct), and treatments (residual), each with $m-1$ d.f. But none of these are natural factors of interest in an OofA experiment.

Our approach for creating OofA designs will employ Van Nostrand's use of PWO factors, which emphasize the order of addition, and so are \textit{a priori} a natural set of factors to use. However, we will not restrict ourselves to certain classes of designs as Van Nostrand and Williams did.

Let $N$ be the number of runs in a design. Define the design matrix $\bfrD$ to be an $N \times m$ matrix, each of whose rows contain a permutation of $c_0, c_1, \ldots, c_{m-1}$, where the $k^{th}$ column of $\bfrD$ shows which component is entered at stage $k$. (This definition will be extended when process variables are also considered.) The full design matrix $\bfrD_F$ is the $m! \times m$ matrix of all such permutations. For any $\bfrD$, an $N \times m'$ \emph{PWO design matrix} $\bfrP$ can be generated, where $m'=m(m-1)/2$. In particular, $\bfrD_F$ generates the full PWO design matrix $\bfrP_F$. Now, it is clear that, given $m$ components, any $m-1$ columns of $\bfrD_F$ (the remaining column being so defined) are necessary to generate $\bfrP_F$. The converse is also true, as we now show.

\begin{proposition0}
For $m$ components, all $m'$ columns from the $\bfrP_F$ matrix are necessary to generate the $\bfrD_F$ matrix.
\end{proposition0}

\begin{proof}
The results are obvious for $m=2$. For $m>2$, suppose that at least one of the PWO-factor columns is not needed. Consider the specific case where $\Fcc{0}{1}$ is not needed. (By the symmetry of the problem, we need to show a result only for a specific case.) But then, for example, the remaining PWO-factor columns could not distinguish between the two permutations $c_0<c_1<c_2<\ldots<c_{m-1}$ and $c_1<c_0<c_2<\ldots<c_{m-1}$. $\blacksquare$

\end{proof}

To create and judge these designs, it will be useful to consider the following, which we do in the next three subsecions:
\begin{enumerate}
\item model-based and balanced-based criteria;
\item measures of goodness;
\item algorithms to create designs.
\end{enumerate}

\subsection{General Criteria and Goodness Measures}\label{sec:GenCrit}
A recent review of several criteria for constructing designs may be found in Lekivetz, et al.\ (2015), some of which we will consider. A common method of constructing designs using model-based criteria employs one or more of the ``alphabetic-optimality'' criteria, for example $D$-optimality, which tries to find a candidate design $\bfrX$ that maximizes $\mathbf{|X'X|}$, where $\bfrX$ is the model matrix. The main-effects model matrix $\bfrX$ corresponding to an OofA design with PWO matrix $\bfrP$ is $\bfrX=[\mathbf{1} | \bfrP ]$, where $\mathbf{1}$ is a conformable vector of 1's.

Alternatives to model-based criteria are balanced-based criteria, and for those it is natural to consider orthogonal arrays (OA's). Here, using notation based on Hedayat, et al.\ (1999), an $N \times k$ array (matrix) $\mathbf{A}$ each of whose columns contain $s$ levels is said to be an OA with strength $t$ if every $N \times t$ sub-array of $\mathbf{A}$ contains all possible $t$-tuples the same number of times---this array may be denoted as $OA(N,s^k,t)$. For mixed-level designs, where $k_i$ factors are at $s_i$ levels, the notation $OA(N,s_1^{k_1}s_2^{k_2}\ldots,t)$ may be used, and the definition can be extended in a natural way. In the usual case where each column corresponds to a factor, an array of strength $t$ corresponds to a design of at least Resolution $t+1$.

There are several ways to measure balance relative to an OA for $t=2$. One, used by Yamada and Lin (1999), is a $\chi^2$ measure. Let $n_{kl}(a,b)$ be the number of rows of $\mathbf{A}$ in which level $a$ appears in column $k$ and level $b$ appears in column $l$. Then the $\chi^2$ measure for the $\{k,l\}$ column pair is
\begin{equation}
\chi^2_{kl}(\mathbf{A}) = \sum_{a=1}^{s_k} \sum_{b=1}^{s_l} \frac{\left[n_{kl}(a,b)-N/(s_k s_l) \right]^2}{N/(s_k s_l)},
\end{equation}
where $s_k$ and $s_l$ are the number of levels in their respective columns and $N/(s_k s_l)=N(s_k/N)(s_l/N)$ is the frequency that would occur under an orthogonal array. An overall measure of non-orthogonality is then
\begin{equation}
\chi^2(\mathbf{A}) = \sum_{1 \leq k < l \leq p}  \chi^2_{kl}(\mathbf{A}) / [p(p-1)/2],
\end{equation}
the average of all such measures, where $\chi^2(\mathbf{A})=0$ for an orthogonal array and $\chi^2(\mathbf{A})>0$ otherwise. The maximum of all such $\chi^2_{kl}(\mathbf{A})$ measures may also be considered.

\subsection{Criteria and Goodness Measures for OofA Experiments}\label{sec:GenCritOofA}
We will examine the model-based approach using the D-criterion, but we first note that it can be limiting for very-small-$N$ designs. For example, Williams' 10-run balanced design in 5 components corresponds to a model matrix $\bfrX$ that is $10 \times 11$, so $\mathbf{|X'X|}=0$. One way to handle this problem would be use the approach suggested by Wu (1993) for supersaturated designs; however, we will not examine very-small-$N$ designs in this paper using the D-criterion.

\begin{figure}[tbp]
\begin{center}
  \scalebox{0.45}{\includegraphics{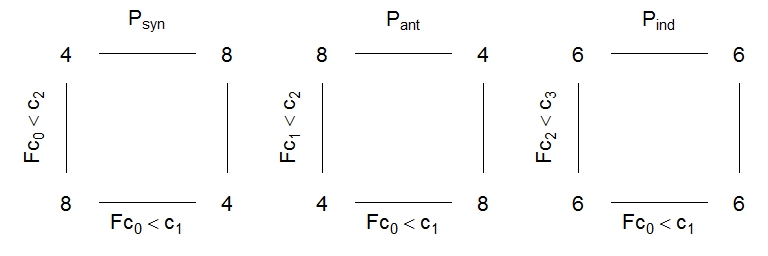}}
  \end{center}
\caption{The three table types of pairs of PWO factors for the $m=4, t=2$ case, with factor examples and frequency counts.}
\label{fig:m4t2tables}
\end{figure}

Next, consider a $\chi^2$, balanced-based, measure of non-orthogonality. It turns out that the $\chi^2$ of Yamada and Lin cannot be used directly for OofA experiments. To see this, consider the $m=4$ case. There are $4!=24$ permutations in the full design, with ${4\choose 2}=6$ PWO factors, as shown in Table~\ref{tab:OofA4_24}, so there are ${6\choose 2}=15$ tables of pairs of PWO factors. Three distinct tables exist and are shown in Figure~\ref{fig:m4t2tables}, along with examples of PWO factors that generated them. In the first table (8 such tables exist), ``$c_0<$'' appears in both PWO factors---this is a \emph{synergistic pair} of PWO factors and is denoted in the table as $P_{syn}$. In the second table (4 such exist), ``$c_1<$'' appears in one factor and ``$<c_1$'' in the other factor---this creates an \emph{antagonistic pair} of PWO factors, denoted as $P_{ant}$. In the third table (3 such exist), one pair of components appears in one factor and the other pair in the other factor---this creates an \emph{independent pair} of PWO factors, denoted as $P_{ind}$. Because there is a natural lack of balance among a subset of these pairs of PWO factors even when \textit{all} combinations are run, the criterion of Yamada and Lin cannot be used directly as a measure of goodness.

Also, note that the $P_{syn}$ and $P_{ant}$ tables, while different, have the same frequency patterns. We call two tables weakly-table-isomorphic (or \emph{wt-isomorphic}) if they have the same frequency patterns; here, in both tables, ``4'' occurs with a frequency of 2 and ``8'' occurs with a frequency of 2. We can use this idea to weakly summarize a design by (a) the number of different wt-isomorphic tables it contains and (b) the frequency of such tables. If we number the tables in our example from least to most balanced and call each one a type of table, then in our example we have two table types, whose corresponding frequencies are 12 and 3.

We call two tables table-isomorphic (or \emph{t-isomorphic}) if one can be obtained from the other by possibly switching one or more labels and/or the directions of inequalities in one of the tables. If two tables are t-isomorphic then they are also wt-isomorphic.

The idea of wt-isomorphism is especially useful when we later consider fractions of these designs. If two such designs are to be compared, we can say that the designs are wt-isomorphic if they have the same frequencies of table types.

Most strongly, we can say that two designs are design-isomorphic (or \emph{d-isomorphic}) if one can be obtained from the other by one or more of these operations:
\begin{enumerate}
\item Permutation of the runs;
\item Permutation of the components;
\item Reversing the order of all orderings in the design, which corresponds to swapping all 0's and 1's in the PWO design matrix.
\end{enumerate}

If two designs are not wt-isomorphic, then they are clearly not d-isomorphic. We will use this idea as a check to see if two designs are not d-isomorphic.

\begin{figure}[tbp]
\begin{center}
  \scalebox{0.45}{\includegraphics{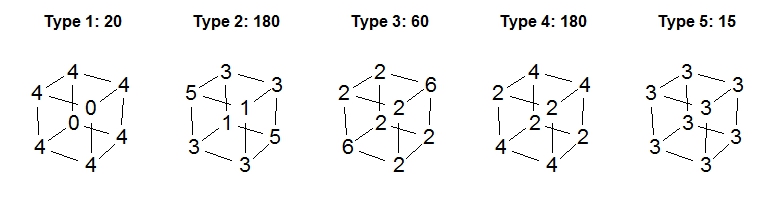}}
  \end{center}
\caption{Five non-wt-isomorphic table types for the $m=6, t=3$ case, with relative (of $N$=24) frequencies for each type. Tables are arranged from least to most balanced.}
\label{fig:m6t3tables}
\end{figure}

For a more complex example, consider $m=6$, with 720 permutations and 15 PWO factors, and strength $t=3$. (The $m=6$ case corresponds to the smallest $m$ that illustrates all the features we show below for $t=3$.) To do this, we need consider the 455 sets of 3 of these 15 factors; these sets of 3 may be represented as $2 \times 2 \times 2$ tables. There are 5 non-wt-isomorphic tables in this case. See Figure~\ref{fig:m6t3tables}, where the tables are arranged from least to most balanced, the PWO-factor labels have been removed, the frequencies in each table have been divided by 30 for ease of understanding, and the frequencies of such tables (out of 455) are shown. Type 1 tables correspond to 3 components, such as $\Fcc{0}{1}, \Fcc{0}{2}, \Fcc{1}{2}$; Type 2 tables to four components, with two components in two PWO's, such as $\Fcc{0}{1}, \Fcc{0}{2}, \Fcc{1}{3}$; Type 3 tables to four components, with one in all three PWO's, such as $\Fcc{0}{1}, \Fcc{0}{2}, \Fcc{0}{3}$; Type 4 tables to five components, with one in two PWO's, such as $\Fcc{0}{1}, \Fcc{0}{2}, \Fcc{3}{4}$;  and Type 5 tables to six components, such as $\Fcc{0}{1}, \Fcc{2}{3}, \Fcc{4}{5}$. As $m$ increases, the relative frequencies of table Types differ---for example, the fraction of tables of Type 5 increases to a limit of $1$.

The natural lack of balance seen in Figure~\ref{fig:m4t2tables}  and Figure~\ref{fig:m6t3tables} leads to the following definition, which includes the PWO-factor case as well as the possibility of additional, process, factors:

\begin{definition0}
 An $N \times m'$ array whose columns consist of all of the PWO factor levels for $m$ components is said to be an Order-of-Addition Orthogonal Array ($\OofAOA$) of strength $t$ for $m$ components if every $N \times t$ sub-array of the array contains all possible $t$-tuples in the relative fractions corresponding to those found when all $m!$ runs are considered. Such an array will be denoted as $\OofAOA(N,m,t)$. If the array is augmented with $p$ additional columns associated with factors whose levels in each column may be varied independently of the first $m'$ columns and each other, and if $N^*_p$ is the number of combinations of these $p$ additional columns, then the resulting $m!N^*_p \times (m' + p)$ array replaces the original $m! \times m$ array.
 \end{definition0}

In particular, the rules for standard OA designs of strength $t$ apply to any $t$-tuple sub-arrays that are created from the process variables alone.

Using this definition and examining Figure~\ref{fig:m4t2tables} and Figure~\ref{fig:m6t3tables}, we can state the following.
\begin{proposition0}
An OofA-OA of strength $2$ exists for $m>3$ only for $N=0 \bmod 12$. An OofA-OA of strength $3$ exists for $m>3$ only for $N=0 \bmod 24$.
\end{proposition0}
In particular, for $m=4$ only the full design with $N=24$ is strength $3$.

We will use these restrictions on $N$ in our search for OA's. These restrictions seem to us to be stronger than for the OA case, and also show that most of Williams designs cannot be OofA-OA's. (For $m=4$, Williams(1949) did provide one design with $N=m(m-1)=12$ runs, but it was not an OofA-OA, and Williams(1950) noted that no solutions exist for those designs for the $m=4$ case, for which $N$ would also have been 12. The next larger $m$ for which $N=0 \bmod 12$ is $m=9$.)

With this definition, we can now generalize the Yamada and Lin measure. For simplicity, we show the result only for the pure OofA case. For $t=2$,  the $\chi^2$ measure for the $\{k,l\}$ column pair for an array $\bfrA$  becomes
\begin{equation}
\chi^2_{kl}(\bfrA) = \sum_{a=1}^{s_k} \sum_{b=1}^{s_l} \frac{\left[n_{kl}(a,b)-NE_{kl}(a,b)/m!) \right]^2}{NE_{kl}(a,b)/m!},
\end{equation}
where $E_{kl}(a,b)$ is the frequency of the occurrence of levels $a$ and $b$ in columns $k$ and $l$ respectively in the full design $\bfrP_F$---see Figure~\ref{fig:m4t2tables} and, after multiplying by 30, Figure~\ref{fig:m6t3tables} for examples of $E_{kl}(a,b)$ values. When $E_{kl}(a,b)=0$, we define that contribution to the sum to be 0. When $p$, say single-d.f., terms for process variables with a possible $N^*_p$ treatment combinations are included, we simply modify $E_{kl}(a,b)$ to be based on the full $m!N^*_p \times (m+p)$ design.

The overall measure of non-orthogonality is again the average of all such measures, which we denote by $\chi^2_{ave,2}(\mathbf{A})$, where the subscript indicates strength 2; we denote the maximum of all such measures as $\chi^2_{max,2}(\mathbf{A})$. A third measure we consider is the fraction of all $\chi^2_{kl}(\bfrA)$ that are 0, which we denote by $FO_2(\bfrA)$.

We will find that in many cases we obtain more than one OofA-OA array for $t=2$, in which case it is natural to consider secondary measures of goodness, in particular the $t=3$ properties of these designs. To this end, we consider all 3-tuples of columns, and so modify $\chi^2_{kl}(\bfrA)$ to, say, $\chi^2_{klu}(\bfrA)$ and call the resulting overall average measure as  $\chi^2_{ave,3}(\mathbf{A})$. Similarly, we modify $FO_2$ to $FO_3$. We also consider $t=2$ or $t=3$ properties when we leave out one component at a time, a projective property of the design. For designs that are not OofA-OA's, we examine $m$ leave-one-out designs and consider the average-over-$m$ measures $\chi^2_{ave,2}$ and $FO_{2}$, which we denote by $\chi^2_{ave,2,-1}$ and $FO_{2,-1}$, respectively. For OofA-OA's, we instead consider $\chi^2_{ave,3,-1}$ and $FO_{3,-1}$.

The above measures all have clear statistical bases. Another measure we consider is the minimum-moment-aberration criterion of Xu (2003). Let $\delta_{ij}(\bfrP)=\sum_{k=1}^m \delta(P_{ik},P_{jk})$ be the number of coincidences in the $i$th and $j$th rows of $\bfrP$, where $\delta$ is the Kronecker delta function. (The Hamming distance between the rows, $m-\delta_{ij}(\bfrP)$, was used by Clark and Dean (2001) to show the equivalence (d-isomorphism idea) of fractional factorial designs.) Xu defines the $s$th power moment of $\bfrP$ to be $K_s(\bfrP)=[N(N-1)]^{-1}\sum_{1\leq i<j\leq N}[\delta_{ij}(\bfrP)]^s$. We standardize this measure of similarity among rows as $Sim_s(\bfrP)=(K_s(\bfrP))^{1/s}$, where $Sim_1\leq Sim_2\leq Sim_3\ldots$ by Jensen's inequality. Designs with the lowest $Sim_1$ values are best---ties are broken by successively considering $Sim_2, Sim_3, \ldots$.

Our emphasis will be in constructing orthogonal OofA designs of strength 2. Efficient algorithms exist for finding OA designs in the case where all columns of the design matrix may be considered independently, e.g. Xu (2002) and Lekivetz, et al.\ (2015), but this is not the case for OofA designs. Efficient algorithms do exist, however, for finding $D$-optimal designs for the OofA case, in which we have a set of candidate points from $\bfrX_F$, from which a subset is selected for the design $\bfrX$. For this reason we consider only the $D$-criterion for most of the design construction below. Our use of this criteria is also partly justified by the following theorem.

\begin{theorem0}Assume that the matrix $\bfrX=[\mathbf{1} | \bfrP ]$ is of full column rank. Then $\chi^2_2=0$ for $\bfrP$ only if $\bfrX$ has a $D$-efficiency of 1 relative to $\bfrX_F$.
\end{theorem0}

\begin{proof}
  If $\chi^2_2(\bfrP)=0$, then $\bfrP$ corresponds to an OofA-OA. In particular, the relative (i.e., $N$-adjusted) frequency of each of the $2 \times 2$ tables that are formed from all pairs of columns of $\bfrP$ correspond to those of $\bfrP_F$. This also applies to the corresponding matrices $\bfrX$ and $\bfrX_F$ because the constant column is orthogonal to all other columns. But these relative frequencies also determine the relative values of the $\bfrX'\bfrX$ matrix, which, by the same reasoning, are equal to those of the $\bfrX_F'\bfrX_F$ matrix. Because of this, the $D$-efficiency of $\bfrX$ with respect to $\bfrX_F$ is equal to 1. $\blacksquare$

\end{proof}

To our surprise, the converse is not true, as we illustrate with a counter-example in Section \ref{sec:OofAOA_24_6_2}; however, for the vast majority of the designs we have created, when $\bfrX$ has a $D$-efficiency of 1 then the corresponding $\chi^2_2=0$. We have also not been able to prove that whenever $\bfrX$ has this $D$-efficiency of 1, then $\bfrX$ is $D$-optimal. However, all of our work suggests this, as does the fact that $\bfrX_F$, the OofA analog of a full factorial, intuitively appears to be the best arrangement in $m!$ runs.

Because $\bfrX_F$ is symmetric in all $m$ components, and because of the similarity of $\bfrX$ to $\bfrX_F$ as noted in the proof above, these OofA designs have properties corresponding to the full design. In particular:
\begin{enumerate}
\item The VIF's for all ${m\choose 2}$ effects in the main-effects model are equal. The VIF value appears to be $3(m-1)/(m+1)$, based on an examination of $m=3,\ldots,8$;
\item The variances of all effects are equal, with the standardized variance of an effect, say $N var \left(\hat{\beta}\right)/\sigma^2$, equal to four times the VIF (a result of the $\left\{0,1\right\}$ coding), where $sigma^2$ is the error variance associated with the linear model corresponding to $\bfrX$;
\item The variances of all predicted values are equal, with value $\sigma^2 \left({m\choose 2}+1\right)/N$.
\end{enumerate}

We use \textit{PROC OPTEX} from SAS/QC software with various numbers of iterations and the Federov option to obtain efficient designs. It is possible to use this procedure to extract all of the designs found or only all of the most efficient designs. We also use the function \textit{optFederov} from the \textit{R} package \textit{AlgDesign} (Wheeler, 2014), but we are only able to extract one design from each function call regardless of the number of iterations. For this reason, our results are often based on a number of designs from SAS software and one from \textit{R} software.

In the next three sections, we present our results. First, we briefly compare our designs to some of the designs created by Van Nostrand and Williams. We then consider designs where OofA-OA's may (and if fact, often do) exist. Our approach usually generates a number of OofA-OA's, and this leads us to consider ways to distinguish among them. Finally, we consider several examples of OofA-OA's in which process-variable factors are also included. Our objective is not to create a catalog of such designs, but rather to illustrate several of them and their properties.

\ssection{RESULTS: COMPARISONS TO PAST WORK}\label{sec:Results}

So that designs in this section may be shown concisely, we will define them through the row-index values of $\bfrD_F$, where the rows of this design are arranged in lexicographic order. (The function \textit{permutations} from the \textit{R} package \textit{gtools} (Warnes, et al., 2015) provides designs in this order.)

Van Nostrand considered a 5-component problem, and using his methods found a 15-run design that corresponds to rows (2, 18, 27, 35, 42, 44, 52, 53, 55, 72, 81, 89, 101, 103, 110) of $\bfrD_F$. The mean \textit{VIF}, used by Van Nostrand, was 3.28. Using our measures for this design, $\chi^2_{ave,2}=1.41$, $\chi^2_{max,2}=5.4$, and $\Deff =0.79$, where $\Deff$ is the $D$-efficiency with respect to $\bfrD_F$.

Our OofA-OA search using the $D$-criterion yielded 13 15-run designs that had $\chi^2_{ave,2}=0.29$, $\chi^2_{max,2}=0.4$, and $\Deff=0.96$, with a mean\textit{VIF} of 2.17, which compares well to the corresponding value of 2 for the full, 120-run, design. One such design had row indices (1, 6, 15, 19, 22, 46, 55, 68, 70, 76, 81, 83, 94, 95, 104). Somewhat surprisingly, none of our 13 designs were wt-isomorphic---this is a consideration we will examine in more detail when we compare OofA-OA's below. Other, secondary measures also showed our design to be better. For example, Van Nostrand's design had $\chi^2_{ave,2,-1}=1.44$ and $Sim_1=5.16$, while the corresponding values for our design were 0.31 and 5.02 (where $Sim_1=5$ for the full design).

We next consider two 5-component Williams' designs. The first, in 10 runs, is balanced for the effects of the preceding treatment,  with row indices (1, 6, 15, 19, 22, 46, 55, 68, 70, 76, 81, 83, 94, 95, 104). Because an $\bfrX$ associated with $m=5, N=10$ cannot be of full-column rank, the $D$-efficiency measure is useless. So we wrote a simple search algorithm using the $\chi^2_{ave,2}$ criterion and found a good design with row indices (3, 10, 32, 38, 46, 64, 86, 94, 99, 101). Williams' design yields $\chi^2_{ave,2}=0.80$ and $\chi^2_{max,2}=3.2$, while ours yields $\chi^2_{ave,2}=0.50$ and $\chi^2_{max,2}=1.7$. We also decided to compare the projective properties of the designs when each component in turn is left out of the design. First consider the average $D$-efficiency of the 5 leave-one-out designs. For the Williams design, $\bfrX$ has column rank of only 6, so no leave-one-out designs have full column rank---the average $\Deff$ is 0. For our design, $\bfrX$ has column rank of 10, and all leave-one-out designs have full column rank, with an average $\Deff$ of 0.84. For Williams design, $\chi^2_{ave,2,-1}=0.88$; for ours, $\chi^2_{ave,2,-1}=0.51$. However, because of the balance features in the Williams design, we find $Sim_1=5$, which is optimal, while our design has $Sim_1=5.02$.

The second 5-component William's design we consider, in 20 runs, is balanced for any number of preceding treatments, with row indices (4, 7, 18, 21, 27, 35, 40, 44, 50, 60, 61, 71, 77, 81, 86, 94, 100, 103, 114, 117). We searched for designs using both $\chi^2_{ave,2}$ and $D$-efficiency criteria, and obtained the following results. Williams' design yields $\chi^2_{ave,2}=0.71$, $\chi^2_{max,2}=1.6$, and $\Deff=0.78$, while ours, using the minimum $\chi^2_{ave,2}$ criterion, yields $\chi^2_{ave,2}=0.15$, $\chi^2_{max,2}=0.8$, and $\Deff=0.90$,  with row indices (2, 9, 20, 28, 36, 37, 42, 51, 52, 56, 72, 78, 81, 83, 89, 101, 103, 109, 112, 116); using the maximum $\Deff$ criterion yields $\chi^2_{ave,2}=0.27$, $\chi^2_{max,2}=1.2$, and $\Deff=0.97$,  with row indices (4, 12, 14, 16, 29, 34, 37, 47, 50, 59, 62, 63, 82, 92, 96, 99, 105, 108, 115, 119). As a secondary measure, the $Sim_1$ values for these three designs are 5, 5, and 5.02, respectively.

So, based on these criteria, designs that are specifically created for OofA considerations lead to better designs than those designed for balance with respect to certain residual-treatment effects.

From this example,we can also see that the connection between OofA-OA's ($\chi^2_{ave,2}=0$) and $\Deff=1$ does not imply that, in comparing two designs, a design with a lower $\chi^2_{ave,2}$ must have a higher $D$-efficiency.

\ssection{RESULTS: SOME OofA OA OR NEAR-OA DESIGNS}

In this section, we consider five OofA designs. Because it turns out that OofA OA's may be constructed for all but the last of them, the notation $\OofAOA (N,m,t)$ will be used to describe the first four.

\subsection{$\OofAOA(12,4,2)$ Designs}
For $\OofAOA(12,4,2)$, 100 starting points from PROC OPTEX produced 88 OA designs, and \textit{AlgDesign} another 1. However, these yielded only two non-wt-isomorphic designs, shown in Table~\ref{tab:OofA4_12}. Both designs had $Sim_1=3$ and $Sim_2=3.34$, the values for the full 24-run design; however, $Sim_3(\bfrP_1)=3.55<Sim_3(\bfrP_2)=3.57$, so Design 1 is superior based on this measure. Design 1 also had superior secondary measures: $\chi^2_{ave,3}=0.82$ vs.\ $1.49$ and $FO_3=0.40$ vs.\ $0.30$.  We were also surprised to see that while the best design had each component balanced with respect to its order of addition, as one might hope for, the second design was quite imbalanced: for example, component 1(2) appeared second in order 0(6) times, and third in order 6(0) times---we will see imbalance in later designs as well. Based on all of these criteria, Design 1 is superior.

\begin{table}[!htbp] \centering
\caption{The two non-wt-isomorphic $\OofAOA(12,4,2)$ designs, with reference-design rows.}
\label{tab:OofA4_12}

\begin{center}
\rule[0pt]{0.3\textwidth}{.4pt}

\rule[18pt]{0.3\textwidth}{.4pt} 

\vspace{-50pt}
\end{center}

\small
\begin{tabular}{@{\extracolsep{-10pt}} cccccr}
\\
\multicolumn{6}{c}{Design 1}\\
  \cline{1-6}
$0$ & $1$ & $3$ & $2$ &   & $2$ \\
$0$ & $2$ & $1$ & $3$ &   & $3$ \\
$0$ & $3$ & $1$ & $2$ &   & $5$ \\
$1$ & $0$ & $2$ & $3$ &   & $7$ \\
$1$ & $2$ & $3$ & $0$ &   & $10$ \\
$1$ & $3$ & $2$ & $0$ &   & $12$ \\
$2$ & $0$ & $3$ & $1$ &   & $14$ \\
$2$ & $1$ & $0$ & $3$ &   & $15$ \\
$2$ & $3$ & $0$ & $1$ &   & $17$ \\
$3$ & $0$ & $2$ & $1$ &   & $20$ \\
$3$ & $1$ & $0$ & $2$ &   & $21$ \\
$3$ & $2$ & $1$ & $0$ &   & $24$ \\
\end{tabular}
\hspace{.25in}
\small
\begin{tabular}{@{\extracolsep{-10pt}} cccccr}
\\
\multicolumn{6}{c}{Design 2}\\
  \cline{1-6}
$0$ & $2$ & $1$ & $3$ &   & $3$ \\
$0$ & $2$ & $3$ & $1$ &   & $4$ \\
$0$ & $3$ & $1$ & $2$ &   & $5$ \\
$1$ & $0$ & $3$ & $2$ &   & $8$ \\
$1$ & $2$ & $0$ & $3$ &   & $9$ \\
$1$ & $2$ & $3$ & $0$ &   & $10$ \\
$1$ & $3$ & $0$ & $2$ &   & $11$ \\
$2$ & $0$ & $1$ & $3$ &   & $13$ \\
$2$ & $3$ & $1$ & $0$ &   & $18$ \\
$3$ & $0$ & $1$ & $2$ &   & $19$ \\
$3$ & $2$ & $0$ & $1$ &   & $23$ \\
$3$ & $2$ & $1$ & $0$ &   & $24$ \\
\end{tabular}

\begin{center}
\rule{0.3\textwidth}{.4pt}
\end{center}

\end{table}

\subsection{$\OofAOA(12,5,2)$ Designs}
For $\OofAOA(12,5,2)$, 100 starting points from PROC OPTEX produced 40 OA designs, and \textit{AlgDesign} another 1. However, all designs were wt-isomorphic. One of these is shown in Table~\ref{tab:OofA5_12}, with secondary measures $\chi^2_{ave,3}=1.24$ and $FO_3=0.42$.

\begin{table}[!htbp] \centering
\caption{The $\OofAOA(12,5,2)$ design, with reference-design rows.}
\label{tab:OofA5_12}
\small
\begin{tabular}{@{\extracolsep{-10pt}} ccccccr}
\\
\hline
\hline
$0$ & $4$ & $2$ & $1$ & $3$ &   & $21$ \\
$0$ & $4$ & $3$ & $1$ & $2$ &   & $23$ \\
$1$ & $0$ & $3$ & $2$ & $4$ &   & $27$ \\
$1$ & $2$ & $3$ & $0$ & $4$ &   & $33$ \\
$1$ & $4$ & $0$ & $2$ & $3$ &   & $43$ \\
$1$ & $4$ & $3$ & $2$ & $0$ &   & $48$ \\
$2$ & $0$ & $3$ & $1$ & $4$ &   & $51$ \\
$2$ & $4$ & $0$ & $1$ & $3$ &   & $67$ \\
$2$ & $4$ & $3$ & $1$ & $0$ &   & $72$ \\
$3$ & $0$ & $2$ & $1$ & $4$ &   & $75$ \\
$3$ & $4$ & $0$ & $1$ & $2$ &   & $91$ \\
$3$ & $4$ & $2$ & $1$ & $0$ &   & $96$ \\
\hline \\
\end{tabular}
\end{table}

\subsection{$\OofAOA(24,5,2)$ Designs}
For $\OofAOA(24,5,2)$, 300 starting points from PROC OPTEX produced 37 OA designs, and \textit{AlgDesign} another 1. An additional 28 OA designs were produced as follows---from the 6 designs discussed below for $\OofAOA(24,6,2)$, each design was used to create 6 $\OofAOA(24,5,2)$ designs by leaving out one component at a time. This created 36 designs, of which 28 were $\OofAOA$'s. From the resulting 66 designs, we were surprised to find that 65 were non-wt-isomorphic---we assign these ID numbers (1--36 OPTEX; 37 \textit{AlgDesign}; 38-65 leave-one-out).

\begin{figure}[tbp]
\centering
  \scalebox{0.60}{\includegraphics{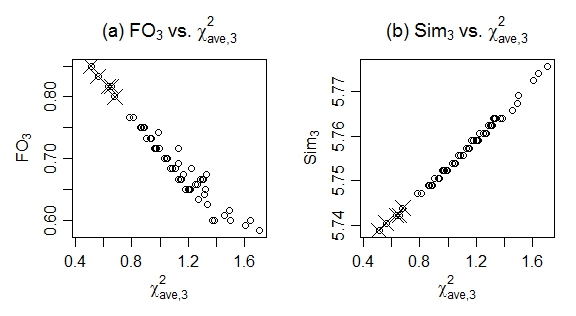}}
  \caption{$FO_3$ and $Sim_3$ vs. $\chi^2_{ave,3}$ for 65 $\OofAOA(24,5,2)$ Designs. The five best designs are shown with overlaid $\times$'s.}
  \label{fig:fMs5_24Chi23VsFOSim3}
\end{figure}

Consider the  $\chi^2_{ave,3}$ measures, which are graphed versus the $FO_3$ and $Sim_3$ measures in Figure~\ref{fig:fMs5_24Chi23VsFOSim3}. (Here, all 65 designs had $Sim_1=5$ and $Sim_2=5.40$, the values for the full design.) Both graphs indicate that five designs, denoted by overlaid $\times$'s, are the best in this group. This is further corroborated by the $\chi^2_{ave,3,-1}$ and the $FO_{3,-1}$ measures---see Table~\ref{tab:OofA5_24Ranks}. These five criteria were then used to rank the 65 designs (tied ranks were assigned to the lowest rank)---these ranks are shown parenthetically in the Table. From these, average ranks ($\overline{\textrm{Rank}}$) were obtained, leading to the ranks (Rank) among these five. The Design ID's (ID) are also provided, as well as the worst measured value among all 65 designs for each of the criteria.

\begin{table}[!htbp] \centering
\caption{Rankings for The $\OofAOA(24,5,2)$ designs.}
\label{tab:OofA5_24Ranks}
\footnotesize
\begin{tabular}{@{\extracolsep{-3pt}} rrrllllll}
\\
\hline
\hline
\\
 ID & Rank & $\overline{\textrm{Rank}}$ & $FO_3$ & $\chi^2_{ave,3}$ & $FO_{3,-1}$ & $\chi^2_{ave,3,-1}$ & $Sim_3$ & $RMV_{ord}$ \\
33 & 1 & 1.0 & 0.85(1) & 0.51(1) & 0.88(1) & 0.43(1) & 5.739(1) & 2.52(59) \\
43 & 2 & 1.6 & 0.83(2) & 0.56(2) & 0.88(1) & 0.43(1) & 5.740(2) & 2.52(59) \\
5 & 3 & 3.4 & 0.82(3) & 0.63(3) & 0.84(4) & 0.58(4) & 5.742(3) & 1.99(36) \\
65 & 4 & 4.2 & 0.80(5) & 0.68(5) & 0.86(3) & 0.51(3) & 5.744(5) & 1.25(1) \\
57 & 5 & 4.8 & 0.82(3) & 0.64(4) & 0.82(7) & 0.65(7) & 5.742(3) & 2.14(42) \\
     &      &  worst:    & 0.58    & 1.71    & 0.63    & 1.45    & 5.776    & 2.86  \\
\hline \\[-1.8ex]
\end{tabular}
\end{table}

However, as for $\OofAOA(24,4,2)$, we next consider the balance of each design with respect to its order of addition, which we now summarize as follows. Define $f_{kl}$ as the frequency with which component $k$ appears in the $l$th order. Then create the unbalance measure $RMV_{ord} = [(1/m^2) \sum_{k=0}^{m-1} \sum_{l=1}^m (f_{kl}-\overline{f_{k.}})^2]^{1/2}$. (Note that full balance ($RMV_{ord}=0$) is not possible for this design because 24 is not divisible by 5.) The $RMV_{ord}$ values in Table~\ref{tab:OofA5_24Ranks} show large differences among the five designs, with the highest-ranked design based on the earlier measures being quite imbalanced in its component-order-of-addition. We believe this large imbalance would be a concern for most experimenters, and the other measures are not too dissimilar, so we would likely recommend design ID 65 in practice. This design, as well as design ID's 5 and 33, appear in Table~\ref{tab:OofA5_24} along with their $f_{kl}$ values.

\begin{table}[!htbp] \centering
\caption{Three $\OofAOA(24,5,2)$ designs (ID's 5, 33, 65) with reference-design rows, followed by $f_{kl}$ values. (For OA columns, see Section~\ref{sec:ProcVar}.)}
\label{tab:OofA5_24}
\begin{center}
\rule{0.7\textwidth}{.4pt}

\rule[18pt]{0.7\textwidth}{.4pt} 

\vspace{-40pt}

\end{center}
\small
\begin{tabular}{@{\extracolsep{-10pt}} ccccccccr} 
 &   &  &  &  &  &  &   &  \\
 &   &  &  &  &  &  &   &  \\
\\[-1.8ex]
& & \multicolumn{7}{c}{Design 5}\\
  \cline{3-9}
$1$ &   & 0 & 1 & 4 & 3 & 2 &   & 6 \\
$2$ &   & 0 & 2 & 1 & 4 & 3 &   & 8 \\
$3$ &   & 0 & 2 & 3 & 4 & 1 &   & 10 \\
$4$ &   & 0 & 3 & 2 & 1 & 4 &   & 15 \\
$5$ &   & 0 & 3 & 4 & 2 & 1 &   & 18 \\
$6$ &   & 1 & 0 & 4 & 2 & 3 &   & 29 \\
$7$ &   & 1 & 2 & 0 & 3 & 4 &   & 31 \\
$8$ &   & 1 & 2 & 4 & 0 & 3 &   & 35 \\
$9$ &   & 1 & 3 & 0 & 2 & 4 &   & 37 \\
$10$ &   & 1 & 3 & 4 & 2 & 0 &   & 42 \\
$11$ &   & 2 & 0 & 4 & 1 & 3 &   & 53 \\
$12$ &   & 2 & 1 & 3 & 4 & 0 &   & 58 \\
$13$ &   & 2 & 3 & 0 & 1 & 4 &   & 61 \\
$14$ &   & 2 & 4 & 3 & 1 & 0 &   & 72 \\
$15$ &   & 3 & 0 & 4 & 1 & 2 &   & 77 \\
$16$ &   & 3 & 1 & 2 & 0 & 4 &   & 81 \\
$17$ &   & 3 & 1 & 4 & 0 & 2 &   & 83 \\
$18$ &   & 3 & 2 & 4 & 0 & 1 &   & 89 \\
$19$ &   & 4 & 0 & 1 & 2 & 3 &   & 97 \\
$20$ &   & 4 & 1 & 0 & 3 & 2 &   & 104 \\
$21$ &   & 4 & 2 & 0 & 3 & 1 &   & 110 \\
$22$ &   & 4 & 2 & 1 & 3 & 0 &   & 112 \\
$23$ &   & 4 & 3 & 0 & 1 & 2 &   & 115 \\
$24$ &   & 4 & 3 & 2 & 1 & 0 &   & 120 \\
\\
$0$ &   & $5$ & $4$ & $6$ & $4$ & $5$ &   &   \\
$1$ &   & $5$ & $5$ & $3$ & $7$ & $4$ &   &   \\
$2$ &   & $4$ & $7$ & $3$ & $5$ & $5$ &   &   \\
$3$ &   & $4$ & $7$ & $3$ & $5$ & $5$ &   &   \\
$4$ &   & $6$ & $1$ & $9$ & $3$ & $5$ &   &   \\
\end{tabular}
\hspace{.15in}
\small
\begin{tabular}{@{\extracolsep{-10pt}} cccccccr} 
 &  &  &  &  &  &   &  \\
 &  &  &  &  &  &   &  \\
\\[-1.8ex]
\multicolumn{8}{c}{Design 33}\\
  \cline{1-8}
 & 0 & 1 & 2 & 4 & 3 &   & 2 \\
 & 0 & 1 & 3 & 4 & 2 &   & 4 \\
 & 0 & 2 & 3 & 1 & 4 &   & 9 \\
 & 0 & 3 & 2 & 4 & 1 &   & 16 \\
 & 0 & 4 & 2 & 1 & 3 &   & 21 \\
 & 0 & 4 & 3 & 1 & 2 &   & 23 \\
 & 1 & 0 & 2 & 3 & 4 &   & 25 \\
 & 1 & 3 & 2 & 4 & 0 &   & 40 \\
 & 1 & 4 & 0 & 3 & 2 &   & 44 \\
 & 1 & 4 & 2 & 3 & 0 &   & 46 \\
 & 2 & 1 & 0 & 4 & 3 &   & 56 \\
 & 2 & 1 & 3 & 0 & 4 &   & 57 \\
 & 2 & 3 & 4 & 0 & 1 &   & 65 \\
 & 2 & 4 & 0 & 1 & 3 &   & 67 \\
 & 2 & 4 & 3 & 1 & 0 &   & 72 \\
 & 3 & 0 & 4 & 1 & 2 &   & 77 \\
 & 3 & 1 & 2 & 0 & 4 &   & 81 \\
 & 3 & 1 & 4 & 0 & 2 &   & 83 \\
 & 3 & 2 & 0 & 1 & 4 &   & 85 \\
 & 3 & 4 & 2 & 1 & 0 &   & 96 \\
 & 4 & 1 & 2 & 0 & 3 &   & 105 \\
 & 4 & 1 & 3 & 0 & 2 &   & 107 \\
 & 4 & 2 & 0 & 3 & 1 &   & 110 \\
 & 4 & 3 & 0 & 2 & 1 &   & 116 \\
\\
 & $6$ & $2$ & $6$ & $6$ & $4$ &   &   \\
 & $4$ & $8$ & $0$ & $8$ & $4$ &   &   \\
 & $5$ & $3$ & $9$ & $1$ & $6$ &   &   \\
 & $5$ & $4$ & $6$ & $4$ & $5$ &   &   \\
 & $4$ & $7$ & $3$ & $5$ & $5$ &   &   \\
\end{tabular}
\hspace{.15in}
\small
\begin{tabular}{@{\extracolsep{-10pt}} cccccccr} 
  &  &  &  &  &  &   &  \\
  &  &  &  &  &  &   &  \\
\\[-1.8ex]
\multicolumn{8}{c}{Design 73}\\
  \cline{1-8}
  & 0 & 1 & 2 & 4 & 3 &   & 2 \\
  & 0 & 2 & 4 & 3 & 1 &   & 12 \\
  & 0 & 3 & 1 & 4 & 2 &   & 14 \\
  & 0 & 4 & 1 & 3 & 2 &   & 20 \\
  & 1 & 0 & 3 & 2 & 4 &   & 27 \\
  & 1 & 0 & 4 & 2 & 3 &   & 29 \\
  & 1 & 2 & 3 & 4 & 0 &   & 34 \\
  & 1 & 3 & 0 & 2 & 4 &   & 37 \\
  & 1 & 4 & 3 & 2 & 0 &   & 48 \\
  & 2 & 0 & 1 & 3 & 4 &   & 49 \\
  & 2 & 0 & 3 & 4 & 1 &   & 52 \\
  & 2 & 1 & 4 & 0 & 3 &   & 59 \\
  & 2 & 3 & 1 & 0 & 4 &   & 63 \\
  & 2 & 4 & 3 & 0 & 1 &   & 71 \\
  & 3 & 0 & 4 & 2 & 1 &   & 78 \\
  & 3 & 1 & 4 & 0 & 2 &   & 83 \\
  & 3 & 2 & 0 & 1 & 4 &   & 85 \\
  & 3 & 2 & 4 & 1 & 0 &   & 90 \\
  & 3 & 4 & 0 & 1 & 2 &   & 91 \\
  & 4 & 0 & 2 & 1 & 3 &   & 99 \\
  & 4 & 0 & 3 & 2 & 1 &   & 102 \\
  & 4 & 1 & 2 & 0 & 3 &   & 105 \\
  & 4 & 2 & 1 & 3 & 0 &   & 112 \\
  & 4 & 3 & 1 & 2 & 0 &   & 118 \\
\\
  & $4$ & $7$ & $3$ & $5$ & $5$ &   &   \\
  & $5$ & $4$ & $6$ & $4$ & $5$ &   &   \\
  & $5$ & $5$ & $3$ & $7$ & $4$ &   &   \\
  & $5$ & $4$ & $6$ & $4$ & $5$ &   &   \\
  & $5$ & $4$ & $6$ & $4$ & $5$ &   &   \\
\end{tabular}
\hspace{.15in}
\small
\begin{tabular}{@{\extracolsep{-10pt}} cccccccc}

\\[-1.8ex]  

\multicolumn{4}{c}{${\scriptstyle OA(2^4)}$} & & \multicolumn{3}{c}{${\scriptstyle OA(2^2 3^1)}$ \rule{0pt}{5.2ex} }\\

$1$ & $2$ & $3$ & $4$ & \hspace{.15in} & $1$ & $2$ & $3$  \\
\hline \\[-2.5ex]
$1$ & $1$ & $1$ & $1$ &   & $1$ & $1$ & $2$ \\
$1$ & $1$ & $1$ & $0$ &   & $1$ & $1$ & $1$ \\
$1$ & $1$ & $1$ & $1$ &   & $1$ & $1$ & $1$ \\
$1$ & $1$ & $1$ & $1$ &   & $1$ & $1$ & $2$ \\
$0$ & $1$ & $1$ & $0$ &   & $0$ & $1$ & $2$ \\
$0$ & $1$ & $1$ & $0$ &   & $0$ & $1$ & $2$ \\
$0$ & $0$ & $1$ & $1$ &   & $0$ & $0$ & $2$ \\
$0$ & $1$ & $0$ & $1$ &   & $0$ & $1$ & $1$ \\
$0$ & $0$ & $1$ & $0$ &   & $0$ & $0$ & $1$ \\
$1$ & $0$ & $0$ & $1$ &   & $1$ & $0$ & $1$ \\
$1$ & $0$ & $0$ & $1$ &   & $1$ & $0$ & $1$ \\
$0$ & $1$ & $0$ & $0$ &   & $0$ & $1$ & $0$ \\
$0$ & $0$ & $1$ & $1$ &   & $0$ & $0$ & $2$ \\
$1$ & $0$ & $0$ & $0$ &   & $1$ & $0$ & $0$ \\
$1$ & $0$ & $1$ & $1$ &   & $1$ & $0$ & $2$ \\
$0$ & $1$ & $0$ & $1$ &   & $0$ & $1$ & $1$ \\
$1$ & $0$ & $0$ & $0$ &   & $1$ & $0$ & $0$ \\
$0$ & $0$ & $0$ & $0$ &   & $0$ & $0$ & $0$ \\
$1$ & $0$ & $1$ & $0$ &   & $1$ & $0$ & $2$ \\
$1$ & $1$ & $0$ & $0$ &   & $1$ & $1$ & $0$ \\
$1$ & $1$ & $0$ & $0$ &   & $1$ & $1$ & $0$ \\
$0$ & $1$ & $0$ & $1$ &   & $0$ & $1$ & $0$ \\
$0$ & $0$ & $0$ & $1$ &   & $0$ & $0$ & $0$ \\
$0$ & $0$ & $1$ & $0$ &   & $0$ & $0$ & $1$ \\
\\
$ $ & $ $ & $ $ & $ $ &   & $ $ & $ $ & $ $ \\
$ $ & $ $ & $ $ & $ $ &   & $ $ & $ $ & $ $ \\
$ $ & $ $ & $ $ & $ $ &   & $ $ & $ $ & $ $ \\
$ $ & $ $ & $ $ & $ $ &   & $ $ & $ $ & $ $ \\
$ $ & $ $ & $ $ & $ $ &   & $ $ & $ $ & $ $ \\
\end{tabular}

\begin{center}
\rule{0.7\textwidth}{.4pt}
\end{center}

\end{table}

\subsection{$\OofAOA(24,6,2)$ Designs} \label{sec:OofAOA_24_6_2}
The $\OofAOA(24,6,2)$ designs were more difficult to obtain, so the number of starting points was increased in PROC OPTEX to 5000. That produced 5 designs with a $\Deff=1$, and \textit{AlgDesign} was used to obtain one more. All designs were non-wt-isomorphic, and were assigned ID numbers (1--5 OPTEX; 6 \textit{AlgDesign}). However, designs 4 and 5 had $\chi^2(\mathbf{A})=0.095>0$, providing the counterexamples noted in Section~\ref{sec:GenCritOofA}.

The four $\OofAOA(24,6,2)$ designs had $Sim_1=7.5$ and $Sim_2=7.96$, the values for the full design, while the other two designs had $Sim_1=7.51$ and $Sim_2=7.97$. More details on all six are provided in Table~\ref{tab:OofA6_24Ranks}. The unbalance measure $RMV_{ord}$ breaks the tie between the top two designs. Although full balance may be possible in this case, none of these designs achieved it. The three highest-ranking designs and the worst design appear in Table~\ref{tab:OofA6_24}, along with their $f_{kl}$ values.

\begin{table}[!htbp] \centering
\caption{Rankings for four $\OofAOA(24,6,2)$ designs and two other $D$-efficient designs.}
\label{tab:OofA6_24Ranks}
\small
\begin{tabular}{@{\extracolsep{-3pt}} rrrllllll}
\\
\hline
\hline
\\
ID & Rank & $\overline{\textrm{Rank}}$ & $FO_3$ & $\chi^2_{ave,3}$ & $FO_{3,-1}$ & $\chi^2_{ave,3,-1}$ & $Sim_3$ & $RMV_{ord}$ \\
6 & 1.5 & 1.6 & 0.69(2) & 1.10(2) & 0.72(1) & 1.00(1) & 8.406(2) & 1.12(1) \\
3 & 1.5 & 1.6 & 0.70(1) & 1.06(1) & 0.70(3) & 1.10(2) & 8.403(1) & 1.79(5) \\
1 & 3 & 3.4 & 0.66(3) & 1.25(5) & 0.70(2) & 1.12(3) & 8.415(4) & 2.01(6) \\
2 & 4 & 3.6 & 0.65(4) & 1.22(3) & 0.67(4) & 1.17(4) & 8.412(3) & 1.74(3) \\
4 & 5 & 4.8 & 0.56(5) & 1.22(4) & 0.57(5) & 1.20(5) & 8.417(5) & 1.53(2) \\
5 & 6 & 6 & 0.52(6) & 1.38(6) & 0.54(6) & 1.36(6) & 8.425(6) & 1.74(3) \\
     &      & worst:  & 0.52    & 1.38    & 0.54    & 1.36    & 8.425    & 2.01    \\
\hline \\
\end{tabular}
\end{table}

\begin{table}[!htbp] \centering
\caption{Three $\OofAOA(24,6,2)$ designs (ID's 6, 3, 1) and another $D$-efficient design (ID 5) with reference-design rows, followed by $f_{kl}$ values.}
\label{tab:OofA6_24}
\begin{center}

\rule{0.8\textwidth}{.4pt}

\rule[18pt]{0.8\textwidth}{.4pt} 

\vspace{-40pt}

\end{center}
\small
\begin{tabular}{@{\extracolsep{-10pt}} cccccccccr}
\\
& & \multicolumn{8}{c}{Design 6}\\
  \cline{3-10}
$1$ &   & 0 & 1 & 5 & 2 & 4 & 3 &   & 20 \\
$2$ &   & 0 & 2 & 4 & 3 & 5 & 1 &   & 40 \\
$3$ &   & 0 & 3 & 1 & 5 & 4 & 2 &   & 54 \\
$4$ &   & 0 & 4 & 5 & 1 & 3 & 2 &   & 92 \\
$5$ &   & 1 & 0 & 3 & 2 & 5 & 4 &   & 128 \\
$6$ &   & 1 & 2 & 3 & 4 & 0 & 5 &   & 153 \\
$7$ &   & 1 & 4 & 3 & 2 & 5 & 0 &   & 208 \\
$8$ &   & 1 & 5 & 3 & 0 & 2 & 4 &   & 229 \\
$9$ &   & 2 & 0 & 5 & 1 & 3 & 4 &   & 259 \\
$10$ &   & 2 & 1 & 4 & 5 & 0 & 3 &   & 281 \\
$11$ &   & 2 & 3 & 1 & 0 & 4 & 5 &   & 295 \\
$12$ &   & 2 & 5 & 0 & 3 & 4 & 1 &   & 340 \\
$13$ &   & 2 & 5 & 4 & 3 & 0 & 1 &   & 359 \\
$14$ &   & 3 & 0 & 4 & 2 & 1 & 5 &   & 375 \\
$15$ &   & 3 & 4 & 5 & 0 & 1 & 2 &   & 451 \\
$16$ &   & 3 & 5 & 2 & 0 & 1 & 4 &   & 469 \\
$17$ &   & 3 & 5 & 2 & 4 & 1 & 0 &   & 474 \\
$18$ &   & 4 & 0 & 2 & 1 & 3 & 5 &   & 487 \\
$19$ &   & 4 & 0 & 5 & 3 & 2 & 1 &   & 504 \\
$20$ &   & 4 & 1 & 5 & 2 & 0 & 3 &   & 525 \\
$21$ &   & 4 & 3 & 1 & 2 & 0 & 5 &   & 561 \\
$22$ &   & 5 & 1 & 0 & 4 & 2 & 3 &   & 629 \\
$23$ &   & 5 & 3 & 1 & 4 & 0 & 2 &   & 683 \\
$24$ &   & 5 & 4 & 2 & 1 & 3 & 0 &   & 712 \\
\\
\\
$0$ &   & $4$ & $5$ & $2$ & $4$ & $6$ & $3$ &   &   \\
$1$ &   & $4$ & $4$ & $4$ & $4$ & $4$ & $4$ &   &   \\
$2$ &   & $5$ & $2$ & $4$ & $6$ & $3$ & $4$ &   &   \\
$3$ &   & $4$ & $4$ & $4$ & $4$ & $4$ & $4$ &   &   \\
$4$ &   & $4$ & $4$ & $4$ & $4$ & $4$ & $4$ &   &   \\
$5$ &   & $3$ & $5$ & $6$ & $2$ & $3$ & $5$ &   &   \\
\end{tabular}
\hspace{.25in}
\small
\begin{tabular}{@{\extracolsep{-10pt}} cccccccccr}
\\
& & \multicolumn{8}{c}{Design 3}\\
  \cline{3-10}
 &   & 0 & 1 & 2 & 5 & 4 & 3 &   & 6 \\
 &   & 0 & 2 & 3 & 4 & 5 & 1 &   & 34 \\
 &   & 0 & 3 & 1 & 4 & 5 & 2 &   & 52 \\
 &   & 0 & 3 & 2 & 5 & 1 & 4 &   & 59 \\
 &   & 0 & 4 & 5 & 1 & 3 & 2 &   & 92 \\
 &   & 1 & 0 & 5 & 2 & 3 & 4 &   & 139 \\
 &   & 1 & 3 & 2 & 4 & 5 & 0 &   & 178 \\
 &   & 1 & 3 & 5 & 0 & 4 & 2 &   & 188 \\
 &   & 1 & 4 & 2 & 5 & 0 & 3 &   & 203 \\
 &   & 1 & 4 & 3 & 0 & 5 & 2 &   & 206 \\
 &   & 2 & 0 & 4 & 1 & 3 & 5 &   & 253 \\
 &   & 2 & 4 & 3 & 1 & 5 & 0 &   & 328 \\
 &   & 2 & 5 & 1 & 3 & 0 & 4 &   & 345 \\
 &   & 3 & 1 & 2 & 0 & 5 & 4 &   & 392 \\
 &   & 3 & 4 & 0 & 2 & 1 & 5 &   & 435 \\
 &   & 3 & 5 & 4 & 1 & 2 & 0 &   & 478 \\
 &   & 4 & 2 & 1 & 0 & 5 & 3 &   & 536 \\
 &   & 4 & 2 & 3 & 0 & 5 & 1 &   & 542 \\
 &   & 4 & 5 & 3 & 1 & 0 & 2 &   & 597 \\
 &   & 5 & 0 & 4 & 3 & 2 & 1 &   & 624 \\
 &   & 5 & 2 & 1 & 4 & 0 & 3 &   & 659 \\
 &   & 5 & 2 & 3 & 0 & 1 & 4 &   & 661 \\
 &   & 5 & 3 & 2 & 4 & 0 & 1 &   & 689 \\
 &   & 5 & 4 & 0 & 1 & 2 & 3 &   & 697 \\
\\
\\
 &   & $5$ & $3$ & $2$ & $6$ & $5$ & $3$ &   &   \\
 &   & $5$ & $2$ & $4$ & $6$ & $3$ & $4$ &   &   \\
 &   & $3$ & $5$ & $6$ & $2$ & $3$ & $5$ &   &   \\
 &   & $3$ & $5$ & $6$ & $2$ & $3$ & $5$ &   &   \\
 &   & $3$ & $6$ & $3$ & $5$ & $2$ & $5$ &   &   \\
 &   & $5$ & $3$ & $3$ & $3$ & $8$ & $2$ &   &   \\
\end{tabular}
\hspace{.25in}
\small
\begin{tabular}{@{\extracolsep{-10pt}} cccccccccr}
\\
& & \multicolumn{8}{c}{Design 1}\\
  \cline{3-10}
 &   & 0 & 3 & 2 & 1 & 4 & 5 &   & 55 \\
 &   & 0 & 3 & 4 & 1 & 5 & 2 &   & 62 \\
 &   & 0 & 4 & 2 & 5 & 3 & 1 &   & 84 \\
 &   & 0 & 5 & 2 & 1 & 4 & 3 &   & 104 \\
 &   & 0 & 5 & 4 & 1 & 3 & 2 &   & 116 \\
 &   & 1 & 2 & 4 & 0 & 5 & 3 &   & 158 \\
 &   & 1 & 3 & 2 & 0 & 4 & 5 &   & 175 \\
 &   & 1 & 4 & 2 & 3 & 5 & 0 &   & 202 \\
 &   & 1 & 5 & 3 & 0 & 4 & 2 &   & 230 \\
 &   & 1 & 5 & 4 & 0 & 3 & 2 &   & 236 \\
 &   & 2 & 1 & 0 & 3 & 5 & 4 &   & 266 \\
 &   & 2 & 3 & 0 & 1 & 5 & 4 &   & 290 \\
 &   & 2 & 4 & 1 & 5 & 3 & 0 &   & 324 \\
 &   & 2 & 5 & 0 & 4 & 3 & 1 &   & 342 \\
 &   & 3 & 2 & 5 & 4 & 0 & 1 &   & 431 \\
 &   & 3 & 4 & 0 & 2 & 5 & 1 &   & 436 \\
 &   & 3 & 5 & 1 & 0 & 2 & 4 &   & 463 \\
 &   & 4 & 1 & 0 & 2 & 3 & 5 &   & 505 \\
 &   & 4 & 3 & 0 & 1 & 5 & 2 &   & 554 \\
 &   & 4 & 3 & 5 & 2 & 1 & 0 &   & 576 \\
 &   & 4 & 5 & 2 & 0 & 1 & 3 &   & 589 \\
 &   & 5 & 0 & 1 & 2 & 3 & 4 &   & 601 \\
 &   & 5 & 2 & 3 & 4 & 1 & 0 &   & 666 \\
 &   & 5 & 3 & 1 & 4 & 2 & 0 &   & 684 \\
\\
\\
 &   & $5$ & $1$ & $6$ & $6$ & $1$ & $5$ &   &   \\
 &   & $5$ & $2$ & $4$ & $6$ & $3$ & $4$ &   &   \\
 &   & $4$ & $3$ & $6$ & $4$ & $2$ & $5$ &   &   \\
 &   & $3$ & $7$ & $2$ & $2$ & $7$ & $3$ &   &   \\
 &   & $4$ & $4$ & $4$ & $4$ & $4$ & $4$ &   &   \\
 &   & $3$ & $7$ & $2$ & $2$ & $7$ & $3$ &   &   \\
\end{tabular}
\hspace{.25in}
\small
\begin{tabular}{@{\extracolsep{-10pt}} cccccccccr}
\\
& & \multicolumn{8}{c}{Design 5}\\
  \cline{3-10}
 &   & 0 & 2 & 4 & 3 & 5 & 1 &   & 40 \\
 &   & 0 & 3 & 1 & 4 & 5 & 2 &   & 52 \\
 &   & 0 & 4 & 2 & 1 & 5 & 3 &   & 80 \\
 &   & 0 & 5 & 1 & 3 & 2 & 4 &   & 99 \\
 &   & 1 & 2 & 0 & 4 & 5 & 3 &   & 148 \\
 &   & 1 & 2 & 3 & 4 & 5 & 0 &   & 154 \\
 &   & 1 & 3 & 0 & 4 & 5 & 2 &   & 172 \\
 &   & 1 & 5 & 4 & 0 & 3 & 2 &   & 236 \\
 &   & 2 & 1 & 0 & 3 & 5 & 4 &   & 266 \\
 &   & 2 & 1 & 4 & 5 & 3 & 0 &   & 282 \\
 &   & 2 & 4 & 0 & 1 & 3 & 5 &   & 313 \\
 &   & 3 & 0 & 2 & 5 & 1 & 4 &   & 371 \\
 &   & 3 & 1 & 2 & 5 & 0 & 4 &   & 395 \\
 &   & 3 & 4 & 0 & 1 & 2 & 5 &   & 433 \\
 &   & 3 & 4 & 2 & 5 & 1 & 0 &   & 450 \\
 &   & 4 & 2 & 0 & 5 & 3 & 1 &   & 534 \\
 &   & 4 & 3 & 1 & 0 & 5 & 2 &   & 560 \\
 &   & 4 & 3 & 5 & 2 & 0 & 1 &   & 575 \\
 &   & 4 & 5 & 1 & 0 & 3 & 2 &   & 584 \\
 &   & 5 & 0 & 1 & 4 & 2 & 3 &   & 605 \\
 &   & 5 & 0 & 2 & 3 & 4 & 1 &   & 610 \\
 &   & 5 & 2 & 3 & 1 & 4 & 0 &   & 664 \\
 &   & 5 & 3 & 2 & 0 & 4 & 1 &   & 686 \\
 &   & 5 & 4 & 1 & 2 & 3 & 0 &   & 706 \\
\\
\\
 &   & $4$ & $3$ & $6$ & $4$ & $2$ & $5$ &   &   \\
 &   & $4$ & $3$ & $6$ & $4$ & $2$ & $5$ &   &   \\
 &   & $3$ & $5$ & $6$ & $2$ & $3$ & $5$ &   &   \\
 &   & $4$ & $5$ & $2$ & $4$ & $6$ & $3$ &   &   \\
 &   & $4$ & $5$ & $3$ & $5$ & $3$ & $4$ &   &   \\
 &   & $5$ & $3$ & $1$ & $5$ & $8$ & $2$ &   &   \\

\end{tabular}

\begin{center}
\rule{0.6\textwidth}{.4pt}
\end{center}
\end{table}

\subsection{Seven-Component Designs}
No $\OofAOA(24,7,2)$ designs were found after 5000 starting points in each of PROC OPTEX and \textit{AlgDesign}. The best design had $\Deff=0.990$, with $\chi^2_{ave,2}=0.07$, with row indices (823, 839, 909, 1167, 1466, 1525, 1653, 1791, 2226, 2258, 2517, 2721, 2927, 2935, 3071, 3515, 3602, 3642, 4001, 4259, 4332, 4415, 4865, 5009). Using the same process, no $\OofAOA(36,7,2)$ designs were found, the best having $\Deff=0.970$ and $\chi^2_{ave,2}=0.29$, with row indices (454, 486, 551, 629, 637, 881, 1296, 1377, 1470, 1529, 1711, 1947, 2068, 2154, 2353, 2382, 2408, 2726, 2794, 2935, 3039, 3117, 3215, 3263, 3340, 3367, 3505, 3649, 3742, 3874, 4060, 4268, 4330, 4559, 4627, 4896); no $\OofAOA(48,7,2)$ designs were found, the best having $\Deff=0.985$ and $\chi^2_{ave,2}=0.22$, with row indices (69, 171, 253, 307, 445, 606, 706, 777, 823, 912, 1009, 1050, 1223, 1547, 1604, 1716, 1756, 1810, 1905, 2021, 2143, 2232, 2284, 2448, 2824, 3030, 3216, 3290, 3357, 3368, 3602, 3806, 3828, 3920, 4013, 4036, 4044, 4182, 4287, 4419, 4463, 4533, 4609, 4754, 4781, 4810, 4842, 4853).

\ssection{RESULTS: COMPONENTS AND PROCESS VARIABLES}\label{sec:ProcVar}

We next investigate the addition of process variables to OofA designs. Formally, we want to consider adding $p$ additional factors to our design, and we let $N^*_p$ be the number of possible combinations of their levels. The levels of these factors may be free to vary independently, as we have suggested to this point for simplicity, but they can, in fact, be constrained---we illustrate both cases below.

Our objective in this section is to illustrate how this may be done, and to this end we consider augmenting a $\OofAOA(24,5,2)$ design with each of the following terms:
\begin{enumerate}
  \item main effects for a $2^2$ design
  \item main effects for a $2^3$ design
  \item main effects for a $2^4$ design
  \item main effects for a $2^{4-1}_{IV}$ (constrained) design
  \item main effects for a $2^2 \times3$ design
\end{enumerate}

Note that the number of treatment combinations for most of these designs divides into 24---such designs were selected with the hope that an $\OofAOA(24,5,2)$ design could be naturally extended. The remaining, $2^4$, design was selected to see whether it, too, might be able to extend an $\OofAOA(24,5,2)$ design.

We could create each design from basic principles---for example, there are $120 \times 16 = 1920$ candidate points for the third design. Our strategy for each design, however, was to first find an $\OofAOA(24,5,2)$ design and use this design instead of the 120-run design. In our example, this reduces the number of candidate points to $24 \times 16 = 384$. In addition, it may also be possible to fractionate the process-variable portion as well---this case leads to the fourth design, with $24 \times 8 = 192$ candidate points.

For our results, we choose to use Design ID 73 as our $\OofAOA(24,5,2)$ design, and we show results only for the $2^4$ and $2^2 \times3$ designs. In obvious notation, we call these designs as $\OofAOA(24,5,2; 2^4)$ and $\OofAOA(24,5,2; 2^2 3^1)$. These results appear in Table~\ref{tab:OofA5_24}, where for brevity we refer to the two process-variable portions as $OA(2^4)$ and $OA(2^2 3^1)$.

The $2^3$ result may be constructed from factors 2--4 from the $2^4$, and the $2^2$ from any two of the factors 2--4. We were not able to find an $\OofAOA$ with the $2^{4-1}_{IV}$ design for Design ID 73, but were able to do so for Design ID 33 (not shown).

\ssection{DESIGNS WITH RESTRICTIONS, AND NOTES ON ANALYSIS}\label{sec:DesWRestrict}

In our work above, we have assumed that all m! permutations are in the set of candidate points. However, this may often not be the case. For example, the experimenter may be convinced that constraints are necessary, such as $c_0<c_1$, or $c_0<c_1$ and $c_2<c_3$, or $c_0<c_1<c_2$. As L. Hare, who has used the Williams' Latin Squares in OofA experiments, notes (personal communication, August 1, 2016), ``I have done some work with order of addition, mostly related to food ingredients. It can make a very big difference for emulsions, for example. I've used a few full permutation designs, but usually the experimenters have some prior knowledge, especially about what orders will not work.'' In such cases, we can simply first reduce the number of candidate points to agree with the restrictions, create a new set of PWO frequency tables of the kind shown in Figure~\ref{fig:m4t2tables}, and use this to define ``orthogonal'' properties in searching for an OofA-OA.

For analyses, we suggest either of two approaches, both of which are based on the assumption that interactions among the PWO factors are likely to occur. The first follows that of Hamada and Wu (1992)---use a stepwise approach to examine active main effects, and then repeat this approach but now including two f.i.'s that are associated with the active main effects. The second is to use tree-based regression modeling, in which interactions of a certain type are examined naturally.

\ssection{FUTURE WORK}\label{sec:FutureWork}

Proving that a $D$-efficiency of 1 is equivalent to $D$-optimality may lead to some insights into the structure of $D$-optimal, and perhaps OofA, designs. We have not been able to create a closed-form method of constructing any OofA-OA's, but have not ruled out such a possibility. For a design with $N$ runs, can the largest $m$ for which an OofA-OA exists be determined for general $N$ and $m$, for a given $t$? A catalog of these designs, including a wide class of component-order-restricted designs and the addition of a wide class of process factors, would be useful, as would be the incorporation of our design-construction methods into software packages that are easily accessible to experimenters. The apparent small number of non-wt-isomorphic designs in some cases and large number in others surprised us. Does a structure exist among these? Can the number of non-wt-isomorphic designs be obtained and cataloged? Graphical summaries of key features of the designs, such as groupings among the orthogonal pairs or triples of the PWO factors, may help experimenters in assigning their components to the component labels $0,\ldots,m-1$. We have relied on the $D$-criterion because of its efficient implementation, but efficient routines for the $\chi^2$-criterion may produce better near-orthogonal OofA designs and designs with better strength-3 properties.

\ssection{CONCLUSIONS}\label{sec:Conclusions}
In this article we have introduced a class of OofA experimental designs, for which designs with good properties may be created in a relatively small numbers of runs. By generalizing the idea of OA's to OofA-OA's, we use a natural set of balance criteria to measure the quality of these designs. A connection made between D-efficient and OofA-OA designs, along with empirical evidence, allows us to use existing D-optimal algorithms to create OofA-OA designs. We have also found necessary conditions for the number of runs needed to create OofA-OA's of strengths 2 and 3. We have created designs for what we believe will be the most common cases, including 12-run OofA-OA's in 4 and 5 components, 24-run OofA-OA's in 5 and 6 components, and 24-run near-OofA-OA's in 7 components. It is possible to extend some of these designs to include process factors, making these designs even more useful. Our methods can be modified to create designs for the common case in which a subset of the component orderings are restricted \textit{a priori} by the experimenter. We also raise a number of questions for future research.

\if0\blind
{
} \fi

\end{document}